\documentclass[12pt]{article}
\usepackage[cp1251]{inputenc}
\usepackage[english]{babel}
\usepackage{amsmath,amsfonts,amssymb}
\usepackage[dvips]{graphicx}

\usepackage{ifthen}
\usepackage{wrapfig}
\usepackage{verbatim}

\makeatletter

 \renewcommand{\@biblabel}[1]{ }         
 \renewcommand{\@makecaption}[2]{%
 \vspace{\abovecaptionskip}
 \sbox{\@tempboxa}{#1. #2}
 \ifdim \wd\@tempboxa >\hsize #1. #2\par
 \else \global\@minipagefalse \hbox to \hsize {\hfil #1. #2\hfil}%
 \fi
 \vspace{\belowcaptionskip} }

\makeatother

\begin{document}

\begin{center}
             \bf\LARGE {Some orbits in various models\\
             of galactic gravitational field}
\end{center}

\begin{center}
  {\large N. V. Raspopova and S. A. Kutuzov   \\}
                            {\small \it Department of Space Technologies and Applied
                            Astrodynamics,   \\
                            Saint Petersburg State University,
                            Russia

                            }
\end{center}

\begin{abstract}
    We consider a gravitational field in steady state
    galaxy models of two kinds. Some of them are axisymmetrical and others are triaxial.
    Equipotentials and potential law are given separately in accordance to Kutuzov and Ossipkov (1980a).
    The relatively simple potential law is based on Kuzmin--Malasidze
    model (1969). Two kinds of models contain four and five
    structural parameters respectively. One composite model is
    suggested as well.
    Some examples of trajectories are calculated in these models.
    The simplest method to describe orbits is drawing their projections
    on coordinate planes. However it needs a great amount of calculation and
    makes troubles in an interpretation of information.
    In the case of axisymmetrical models a motion in co-moving meridional
    plane (with cylindrical coordinates R, z) is considered as a common way.
    In the case of triaxial models one can use three different
    co-moving planes passing through moving star and corresponding coordinate axis.
    We describe models in sections 2--4, calculated orbits are discussed in section
    5.

\end{abstract}

\section{Introduction}

    In recent years trajectories of stars in various models of the
    gravitational field of the galaxies have been constructed intensively.
    We intend to clear up how do orbit characteristics depend on model
    properties.
    Similar investigation was fulfilled by Kutuzov \& Ossipkov (1992), where box
    eccentricities and maximal elevation of 70 orbits of open
    clusters in three models were compared. Here we investigate relation
    of the orbit properties with variation of the properties
    of different in principle models --- axially symmetric and
    triaxial. This work continues our  research (Kutuzov \& Raspopova 2008).

    There is a great variety of models in literature. We mention just few
    of them. Ossipkov (1997)
    has performed an exhaustive analysis of fourth-order
    (quaternary) equipotentials of general form with some parameters for
    systems with rotational symmetry.

    Dynamical models of triaxial elliptical galaxies were constructed
    by Arnold et al. (1994) using St\"{a}ckel potential for
    calculating the dependence velocity field of the model on the
    shapes of orbits.  Merritt and Fridman (1996)
    showed essential role of the stochastic orbits in triaxial
    gravitational field with central density cusps. For
     similar  model Merritt and Valluri (1996) found that
    central point mass  designed to represent nuclear black hole
    causes stochastic orbits to diffuse through phase space.
    Kutuzov (1997) suggested the model with eighth-order
    equipotentials, providing triaxial mass distribution.

    Here we use simpler triaxial fourth-order
    equipotentials of special form (Kutuzov 1998). One of the parameters,
    responsible for the triaxial shape, is specified in the form of a variable in such a way that
    equipotentials asymptotically tend to spheres, thus, the natural requirement for the
    models of finite mass is satisfied. A rarely
    used property, according to which the dependence of the
    coefficients in the equation of level surfaces on the parameters
    of the family does not change the order of the equation, is employed. In the
    special case a biaxial disk embedded in the halo occurs.

\section{Family of axially symmetric models}

    All the quantities are dimensionless. The change to
    dimensional quantities is fulfilled by multiplication ones
    by the corresponding scaling parameters. Basic ones are units of
    length, potential and mass
\begin{equation*}
   \hat r, \quad \hat \Phi, \quad \hat {\cal M} = \hat r \hat \Phi /G ,
\end{equation*}
    where  $ G $ is gravitational constant.

    At first, we consider axisymmetrical models. The potential law
    is specified analytically with free parameters and
    the one independent variable $ \xi $. The only argument $ \xi $ is a
    function of coordinates
\begin{equation}\label {fi}
    \varphi = \varphi (\xi) \,, \quad \xi^2 = f (R,\; z)\,.
\end{equation}
    The variable $ \xi $ is the equipotential parameter that changes from one
    equipotential to another, determining their family. Putting $ \xi = $ const
    formula (\ref{fi}) gives equation of a fixed equipotential.

    Equipotentials and potential
    law are given separately in accordance to Kutuzov \& Ossipkov (1980a).
    Ten models being the special cases of
    mentioned family were listed by Kutuzov \& Ossipkov (1980b).
    These models were suggested earlier by P.~P.~Parenago (1950, 1952), G.~M.~Idlis (1961),
    G.~G.~Kuzmin (1953, 1956), A.~Toomre (1963),
    H.~Plummer (1911), M.~Miyamoto and R.~Nagai (1975), M.~H\'{e}non (1959), G.~G.~Kuzmin
    and \"{U}.-I.~K.~Veltmann (1972). Euipotentials, constructed there, were generalized later
    (Kutuzov 1989). According to
    (\ref{fi}), equipotentials are determined by the function $ f (R,\; z) $.
    Two-parametric family of equipotentials was suggested~(Kutuzov
    1989)as follows
\begin{equation}\label {xi}
    \xi^2 = r^2 + 2 \mu (1 - \varepsilon)(q - \varepsilon)\,,
    \quad  r^2 = R^2 + z^2\,,
    \quad  q = \sqrt {\varepsilon^2 + z^2}\,.
\end{equation}

    Quantities $ \varepsilon,\; \mu $ are structural
    parameters of the model
\begin{equation*}
  0 \leq \varepsilon \leq 1 \; ,  \qquad 0 \leq \mu < \infty \; .
\end{equation*}

    The first term on the right-hand of equation describes the
    spherical shape of equipoten\-tials at infinity. The second term
    causes to a disk component in the model when $ \varepsilon \to 0 $.
    These equipotentials coincide with the ones
    of the model of Miyamoto \& Nagai (1975) for $ \mu = 1
    $. Let us compare our equipotentials with the model of Satoh
    (1980). We perform the dimensional potential of
    that model to
\begin{equation*}
    \Phi(\xi) = \frac{G \cal M}{\hat r\sqrt{1-\varepsilon^2+\xi^2}},
    \quad  \xi^2 = r^2+2(1-\varepsilon)(q-\varepsilon).
\end{equation*}
    It is obvious that these equipotentials are a special case of
    our ones \eqref{xi} when $ \mu=1 $. For $ \varepsilon=1 $ the model is spherical,
    potential has discontinuity in the center, where $ \xi=0 $. System
    degenerates into mass point with  Kepler potential.

    We take another potential law
\begin{equation}\label{pl}
    \varphi(\xi) = \frac {\alpha} {\beta + \sqrt {1 + \varkappa \xi^2}}\, \qquad
    \beta = \alpha - 1\,,
\end{equation}
    that coincides in the plane $ z = 0 $ with Kuzmin--Malasidze law(1969).
    Thus accepted family of the models has
    four structural dimensionless parameters
    $\varepsilon,\:\mu,\:\alpha,\:\varkappa$. It gives great opportunities for
    modeling of the various galaxies and star clusters. The
    potential is normed so that in the center $ \varphi(0) = 1 $.

    Asymptotics of the potential allows us to find the expression for the
    dimensionless mass of the model
\begin{equation*}
               {\cal M} =   \frac {\alpha} {\sqrt{\varkappa}}\,.
\end{equation*}
    When $ \varepsilon = 1 $ or $ \mu = 0 $  the force field is spherically symmetrical,
    because $ \xi = r $ according to (\ref{xi}). In the case of $ \varepsilon = 0 $
    the force has a discontinuity in the $ z = 0 $ plane, as $ q = |z| $. This
    determines the existence of an infinitely thin circular disk,
    imbedded into continuous halo; dimensionless mass of the disk is
    equal $ \mu $ (Kutuzov 1989).

\section{Family of triaxial models}

    We still take potential in the form of (\ref{fi}), but now its
    argument $\xi $ is the function of three Cartesian coordinates
    $ x, y, z $ (with the origin at center of mass)                             \hfil
\begin{equation*}
              \xi^2 = g (x, y, z ) \;.
\end{equation*}
    Clearly, the addition of $\xi$ to the list of arguments, i.e.
    the representation  $ f(x,y,z,\xi) $, does not change the order
    of the equipotential equation relative to the  $ x, y, z $ coordinates.                  \hfil

    Let the function $ f(x, y, z, \xi) $ consist of three terms
\begin{equation}\label{f3}
 \xi^2 = r^2 + 2 \, \mu (1 - \varepsilon) \, (q - \varepsilon) + \tau (\xi) \, s^2 \;,
            \quad   0 \leq \xi < \infty \; .
\end{equation}
         Here,
\begin{equation*}
    r^2 \equiv x^2 + y^2 + z^2 \; ,                          \quad
    q   \equiv \sqrt{\varepsilon^2 + z^2} \; ,
\end{equation*}
\begin{equation*}
    s^2 \equiv y^2 - x^2
    \begin {cases}
      \geq 0, & |x| \leq |y|  \, ;    \\
      <    0, & |x|  >   |y|  \, .
    \end   {cases}
\end{equation*}
    The new expression differs from (\ref{xi}) by his third term, that is responsible for the triaxial shape.
    Coefficient $ \tau $ is assumed to be a function of $ \xi $.
    Let us call it the triaxiality parameter. Sometimes it is
    convenient to express $ \tau $ in terms of some bounded function
    $ \chi(\xi) $ that is proportial to it                            \hfil
\begin{equation}\label{tau}
 \tau (\xi) = (1 - \varepsilon) \, \chi(\xi) \; .
\end{equation}
        For  $ \varepsilon = 1 $  $ r^2 $ remains on the right side of eq.~(\ref{f3}),
    which implies that the model is spherical. But now equality $ \mu = 0 $
    does not mean transfer to the spherical model, because the triaxiality
    parameter might be not equal zero. Below we'll discuss function $\tau (\xi)$ in
    details. We only note here that for $ \tau \equiv 0 $ the model
    ceases to be triaxial, including models with rotational symmetry
    (\ref{xi}) as a special case. Note also that the parameter $ \xi
    $ is expressed in terms of the $ x, y, z $ coordinates
    implicitly. To calculate it from the coordinates, we must solve
    eq.~(\ref{f3}) using eq.~(\ref{tau}), for example, by successive
    approximations by assuming initially that $ \tau = 0 $.

    The equation of equipotential (\ref{f3}) has fourth order
    with respect to coordinates. It gives the second-order equation in
    equatorial plane
\begin{equation}\label{el}
 \xi^2 = (1 - \tau)\, x^2 + (1 + \tau)\, y^2 \; .
\end{equation}
    The section of the equipotential by the equatorial plane is an
    ellipse for $ |\tau| < 1 $. Equation~(\ref{el}) yields the
    following expressions for the semi-major and semi-minor axes of
    the ellipse                                                     \hfil
\begin{equation}\label{a2b2}
        a^2 = \frac{\xi^2}{ 1 - \tau (\xi) } \;, \quad
        b^2 = \frac{\xi^2}{ 1 + \tau (\xi) } \;.
\end{equation}
    If $ 0  \leqslant \tau (\xi) < 1 $, the two axes are real, and
        $ b  \leqslant \xi  \leqslant a $ .

     Kutuzov (1998) suggested the following constraints on the triaxiality parameter $ \tau (\xi)
     $. It is desirable (but not necessary) that, as one approaches
     the system's center, the ellipse \eqref{el} changes to a
     circumference that would shrink to a point
\begin{equation}\label{t00}
               \tau (0) = 0 \; .
\end{equation}
For the potential to become spherically symmetric at infinity, the
following limit must exist
\begin{equation}\label{t01}
               \lim_{\xi \to \infty}  \tau (\xi)  = 0 \; .
\end{equation}
In this case, $\tau$ must tend to zero as some negative power of
$\xi$. The third term on the right side of \eqref{f3} will then have
an order that is less than 2. Since the second term has only the
first order in $z$, the first term, which always has order 2,
dominates.

In order that the semimajor axis $a$ of the ellipse \eqref{el} be no
smaller than its semiminor axis $b$, and that both axes remain
bounded for a finite $\xi$, we assume that
\begin{equation}\label{t02}
               0  \leqslant \tau (\xi) < 1 \; .
\end{equation}

The equipotentials of the family must not intersect (and touch) each
other. This requires that the axes $a$ and $b$ be monotonically
increasing functions of $\xi$
\begin{equation*}
    a^\prime (\xi) > 0 \; , \qquad  b^\prime (\xi) > 0 \; .
\end{equation*}
    The prime denotes differentiation with respect to $ \xi^2 $.
    Taking into account \eqref{a2b2}, we obtain differential
    inequalities which the triaxiality parameter must satisfy
\begin{equation*}
    \xi^2\tau^\prime (\xi) - \tau(\xi) + 1 > 0, \quad \xi^2\tau^\prime (\xi) - \tau(\xi) - 1 <
    0.
\end{equation*}

    Here are three examples of the function $ \chi(\xi) $,
    that satisfy the necessary conditions
\begin{equation}\label{3hi}
           \chi_1(\xi) = \displaystyle \frac{ A_1 }     {1 + \xi^2} \; ,          \quad
           \chi_2(\xi) = \displaystyle \frac{ A_2 \xi^2}{1 + \xi^4} \; ,          \quad
           \chi_3(\xi) = \displaystyle \frac{ A_3 \xi } {1 + \xi^2} \;,
\end{equation}
    with the following constraint on the positive parameters
\begin{equation*}
          A_i < \frac{1}{1 - \varepsilon}, \quad i = 1,2,3 \,.
\end{equation*}
    The first function does not satisfy constraint (\ref{t00}), so that the
    model is also triaxial in the central region. The function $ \tau $ is
    defined in terms of $ \chi $ by formula (\ref{tau}). The coefficients $ A_i $
    allow us to make the triaxiality parameter $ \tau $ arbitrarily
    small.

    We still use expression (\ref{pl}) for the potential law.
    In the case when the coefficients $ A_i $ are equal zero, the model coincides with
    axially symmetric one entirely.

    A circular velocity has no sence in triaxial model. So we introduce
    quasi-curcular velocity, that just  characterizes the
    gravitational field
    \begin{equation}\label{kwC}
    W^2 = - R\: \left[\frac{\partial\psi(R,\;z)}{\partial R}\right]_{z=0}, \quad \psi(R,\;z)=\left.\varphi(\xi)\right|_{x=y}, \quad R=\sqrt{x^2+y^2}.
    \end{equation}

\section{Composite models}

    Two families of the models considered in sections~2,~3 give us
    favorable opportunities for modeling the various galaxies and star
    clusters. But we get more flexible instrument if we use
    super\-position of the models of these families. We define the set
    of parameters as
\begin{equation*}
            p_n = \{\varepsilon_n,\, \mu_n,\, \alpha_n, \,\varkappa_n, A_i^{(n)} \}\,,
            \qquad n=\overline{1, N}.
\end{equation*}
    here $ i $ --- number of the selected expression in  (\ref{3hi}).
    Then N--component model of the gravitational field is the
    weighted sum of the potentials (\ref{pl}), (\ref{f3})
\begin{equation*}
            \varphi = \sum_{n=1}^N w_n\varphi_n(\xi_n)\,,
            \qquad \sum_{n=1}^N w_n = 1\,,
\end{equation*}
    where $ w_n $ are assigned normalized weight coefficients.
    Mass densities are summed up in just the same way. For example, if
    we take
    $ N = 2 $, $ \varepsilon_1 = 0 $, $ \varepsilon_2 > 0 $,
    $ A_k^{(1)} = A_k^{(2)} = 0 $,
    and all the other model parameters in both components we retain unchanged
    then we have five free parameters, including $ w_1 $. In such a
    model force relief field might involve crater either do
    not involve or involve crater with central peak.

\section{Investigation and discussion of orbits}

\textbf{5.0.} We solve Cauchy problem to construct the trajectory of
the star. The system of ordinary
    differential equations of the second order is
\begin{equation*}
    \mathbf{\ddot{r}} = \nabla \varphi (\xi)\,.
\end{equation*}
    Here $\mathbf{r}$ is a position vector of the star with respect to the center.
    Initial conditions are given for arbitrary moment of time $ t_0 $:
\begin{equation*}
  \mathbf{r} (t_0) = \mathbf{r}_0, \qquad \mathbf{v} (t_0) = \mathbf{v}_0,
\end{equation*}
    where ${\bf v} = \dot{\mathbf{r}}$ --- velocity vector of the star.

    We use dimensionless time (and all the other functions and variables).
    Interval of integration is characterized by number of crossings the plane $z=0$ or $y=0$.

    System of ordinary differential equations of the second order is transformed to the
    form of equations of the first order in a usual way and is solved numerically. We used Merson's
    method and Dormand-Prince one with variable time step
    (Hairer, Nersett, Wanner 1990) in order to control specified accuracy.
    We have chosen these methods, as they supplied rather small relative error for the integral of energy.

\textbf{5.1}.
   Energy $ E $ and $z$-component of the angular momentum $ L_z $
   are the integrals of motion for the axisymmetrical model (Ogorodnikov 1958)
\begin{equation*}
    E = \varphi(\xi) - \frac{1}{2}(v_m^2 + v_{\theta}^2)\,, \quad
        L_z = R v_{\theta}\,, \quad v_m^2 = v_R^2 + v_z^2 \,.
\end{equation*}
    Here $ v_R $, $ v_{\theta} $, $ v_z $ are the velocity
    components in cylindrical coordinates, $ v_m $
    is a meridional velocity, $ E $ is the total energy with the opposite sign.
    Using $z$-component of the angular momentum one can write the equations of
    motion as a system of five equations of the first order (Kutuzov \& Ossipkov 1981):
\begin{equation}\label{eqmR}
\left\{ \begin{aligned}
   \dot{R}      & = v_R    \,,& \displaystyle {\dot v}_R & = L_z^2/R^3 + F_R \,,       \cr
   \dot{\theta} & = L_z/R^2\,,&    &   \cr
   \dot{z}      & = v_z    \,,& \displaystyle {\dot v}_z   & = F_z \,,
        \end{aligned} \right.
\end{equation}
$F_R$, $F_z$ are the components of force $\mathbf{F}$ per unit mass.

    We calculate trajectories of the stars either with the
    initial zero full velocity or with the initial zero meridional velocity. That stars we
    call falling ones. Some periodic orbits were found (Fig.~1).
    Fig.~1a shows the orbit of the star,
    falling from the contour of zero-velocity curve.
    Fig.~1b presents the orbit of the falling star, with the initial zero full
    velocity.

\begin{figure}[p]
\centering
\begin{minipage}[t]{0.49\textwidth}
\centering \includegraphics[width=\linewidth]{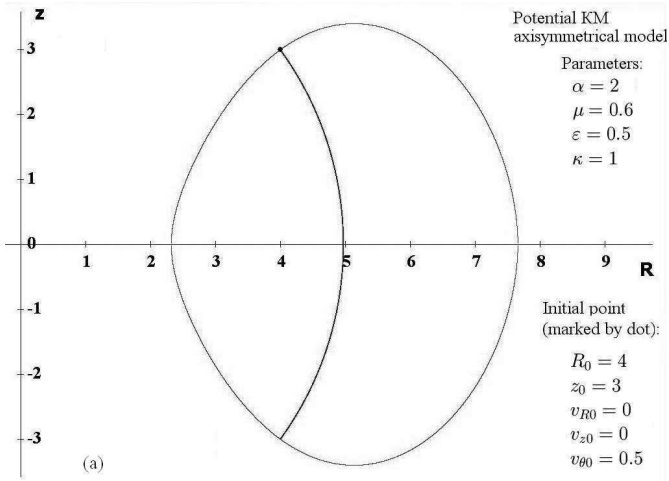}
\end{minipage}\hfill
\begin{minipage}[b]{0.49\linewidth}
\centering \includegraphics[width=\linewidth]{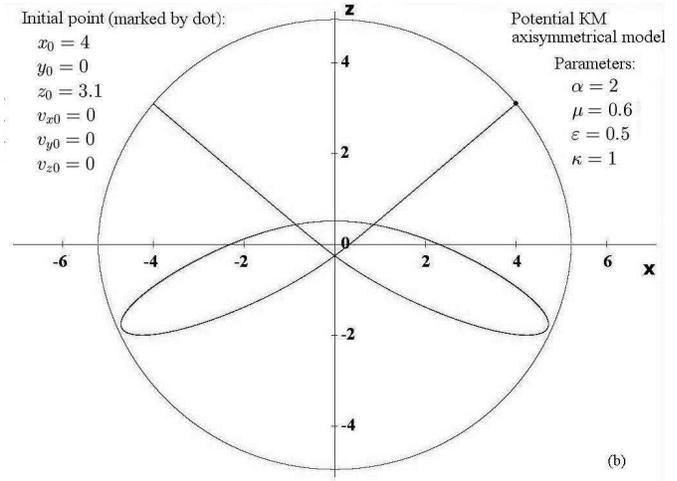}
\end{minipage}
\caption{Periodic orbits in axisymmetrical model: the trajectory of
the falling star inside the area bounded by the zero-velocity
curve.}
\end{figure}

    If we assign small perturbations in the initial
    phase coordinates these orbits conserve the form, filling the
    tube with varying width, which includes initial periodic parent
    orbit~(Fig.~2a). We form perturbations such a way so to make star
    to begin its motion from the equipotential passing through
    the initial point of the parent orbit. As we see in Fig.~2b
    the orbit does not conserve its form and does not include parent orbit
    if perturbations are rather large. We have similar result while varying velocity components.
    Notice that trajectory does not reach
    equipotential curve in the half-plane $z<0$ in  Fig.~1b,~2a .

\begin{figure}[p]
\centering
\begin{minipage}[t]{0.49\textwidth}
\centering \includegraphics[width=\linewidth]{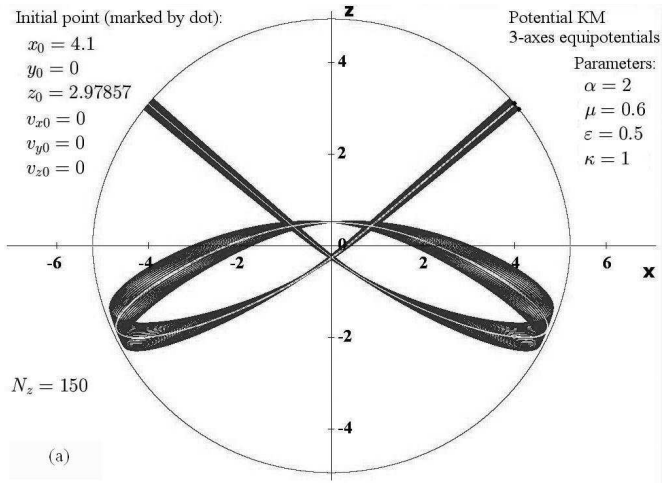}
\end{minipage}\hfill
\begin{minipage}[b]{0.49\linewidth}
\centering \includegraphics[width=\linewidth]{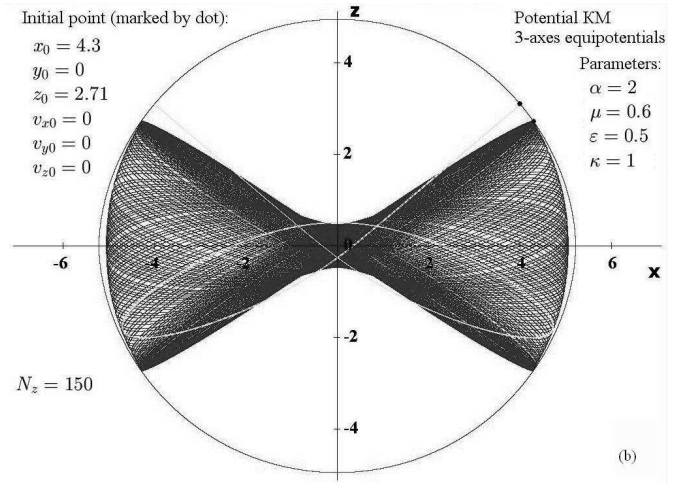}
\end{minipage}
\caption{Effect of varying initial conditions for periodic orbit in
Fig.1b.}
\end{figure}

\begin{figure}[p]
\centering
\begin{minipage}[t]{0.49\textwidth}
\centering \includegraphics[width=\linewidth]{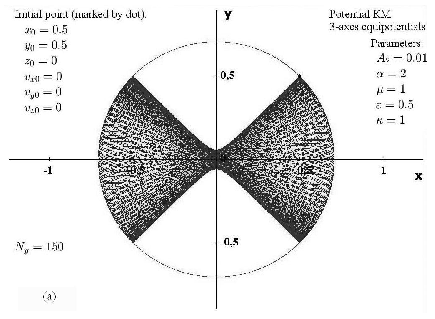}
\end{minipage}\hfill
\begin{minipage}[b]{0.49\linewidth}
\centering \includegraphics[width=\linewidth]{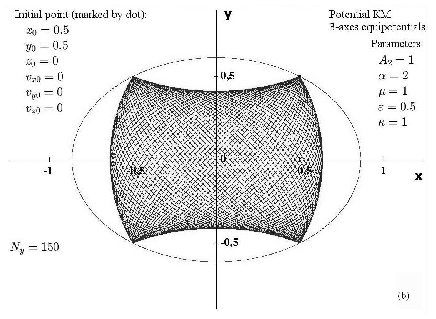}
\end{minipage}
\caption{Variations of the coefficient $A_2$ in triaxility parameter
$\tau$.}
\end{figure}

\begin{figure}[p]
\centering
\begin{minipage}[t]{0.49\textwidth}
\centering \includegraphics[width=\linewidth]{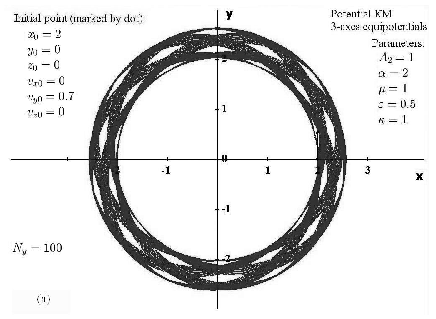}
\end{minipage}\hfill
\begin{minipage}[b]{0.49\linewidth}
\centering \includegraphics[width=\linewidth]{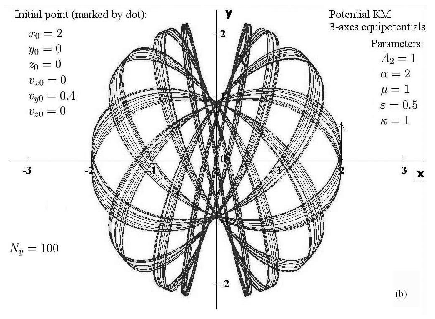}
\end{minipage}
\caption{Loop and box orbits in triaxial model}
\end{figure}

\begin{figure}[h]
\begin{center}
  \includegraphics[width=0.6\textwidth]{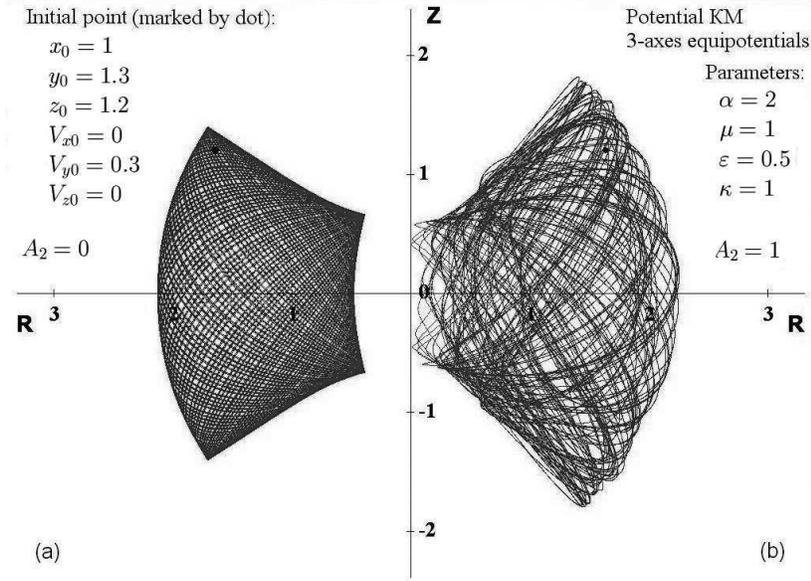}
\caption{Two trajectories in their co-moving planes.  (a) --- in
 axysimmetrical model; (b) --- in triaxial model.
 Iniatial positions and
 velocity components are the same for both trajectories }
\end{center}
\end{figure}

\begin{figure}[h]
\centering
\begin{minipage}[t]{0.49\textwidth}
\centering \includegraphics[width=\linewidth]{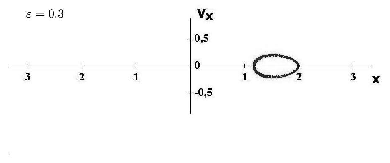}
\end{minipage}\hfill
\begin{minipage}[b]{0.49\linewidth}
\centering \includegraphics[width=\linewidth]{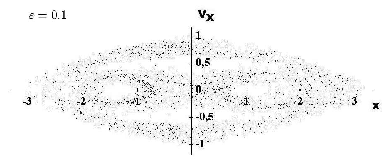}
\end{minipage}
\centering
\begin{minipage}[t]{0.49\textwidth}
\centering \includegraphics[width=\linewidth]{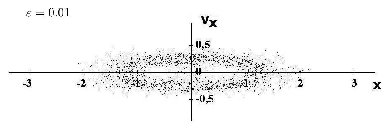}
\end{minipage}\hfill
\begin{minipage}[b]{0.49\linewidth}
\centering \includegraphics[width=\linewidth]{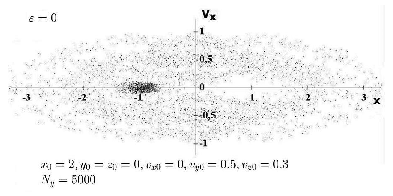}
\end{minipage}
\caption{Surfaces of section for orbits in triaxial models  with
 various flattening.
 Initial positions are the same for all orbits, values
 of parameters are the same too, except  $\varepsilon$. $N_y$ is the
 number of crossing the plane $y=0$ with positive $v_y$. }
\end{figure}

\textbf{5.2}. We consider orbits in triaxial model further.
Trajectories of a star, falling from the same initial point in the
fields with various triaxiality parameters, are shown in Fig.~3.

    It's also interesting to consider trajectory when the inial point lies on the $x$-axis,
    while initial velocity is directed along the $y$-axis.
    For the axially symmetric model the trajectory lies inside the area bounded by
    circular orbit if the initial velocity
    (its nonzero component $v_y$) is smaller than circular velocity
    at the initial point. It lies outside if the initial velocity
    is larger than circular one. The same situation is observed for the triaxial model~(Fig.~4)
    in respect of the quasi-circular velocity (\ref{kwC}).

    It's convenient to study orbits in the co-moving meridional plane for the
    axi\-symmetrical potentials~(Fig.~5a). We construct orbits in the analogical plane
     for the triaxial model~(Fig.~5b).
    Bounds of the orbit become more ``dishevelled''.

    By analogy with Poincar\'e surfaces of section (Binney \& Tremaine 1987)
    we can construct six diagrams. The first pair of them has axes
    $ x,\; \dot x $. Points with these coordinates are plotted at the
    moments when $ y = 0,\; \dot y > 0 $ or $ z = 0,\; \dot z > 0 $
    respectively. Other diagrams could be obtained by cyclic permutation of coordinates.
    Surfaces of section $(x,v_x)$ when $ y = 0,\; v_y > 0 $ for the triaxial orbits are
    shown in~Fig.~6. Calculations are made for the models with various values of
    $\varepsilon$. We remind that the value $\varepsilon=1$ means a
    spherical model and $\varepsilon=0$ means a model with embedded
    sharp disk. Obviously a stochasticity grows with a flattening.

    We plan to continue the calculation and analysis of orbits in
    various orbits, describing real galaxies.

\subsection*{Acknowledgments}

 This work was partly supported by the RFBR (Grant 08-02-00361).

\end{document}